\documentclass[reprint,unsortedaddress,amsmath,amssymb,aps,pra,showpacs]{revtex4-1}
\usepackage{amsmath}%
\usepackage{amsfonts}%
\usepackage{amssymb}%
\usepackage[cal=cm]{mathalfa}
\usepackage{graphicx}
\usepackage{dcolumn}
\usepackage{bm}
\usepackage{braket}
\usepackage{sidecap}
\usepackage{color}
\usepackage{bbm}
\usepackage{nicefrac}
\usepackage{float}
\usepackage{placeins}
\usepackage[dvipsnames]{xcolor}
\usepackage[utf8]{inputenc}
\usepackage[normalem]{ulem}
\usepackage{nicefrac}
\usepackage{dsfont}
\usepackage{xcolor, soul} 
\usepackage[normalem]{ulem} 
\usepackage{siunitx}
\usepackage[citebordercolor=blue]{hyperref}
\usepackage[utf8]{inputenc}
\usepackage{gensymb}
\begin{document}

\title{Twisting Neutral Particles with Electric Fields}

\author{Niels Geerits$^{1}$}
\email{niels.geerits@tuwien.ac.at}
\author{Stephan Sponar$^1$}
\affiliation{%
$^1$Atominstitut, Technische Universit\"at Wien, Stadionallee 2, 1020 Vienna, Austria}

\date{\today}
\pacs{03.65.-w, 03.75.Be, 03.65.Vf}
\hyphenpenalty=800\relax
\exhyphenpenalty=800\relax
\sloppy
\setlength{\parindent}{0pt}

\noindent
\begin{abstract} 
	We demonstrate that spin-orbit coupled states are generated in neutral magnetic spin 1/2 particles travelling through an electric field. The quantization axis of the orbital angular momentum is parallel to the electric field, hence both longitudinal and transverse orbital angular momentum can be created. Furthermore we show that the total angular momentum of the particle is conserved. Finally we propose a neutron optical experiment to measure the transverse effect.
\end{abstract}

\maketitle

\section{Introduction.} Intrinsic orbital angular momentum (OAM) has been observed in free photons \cite{Allen1992,Enk1994,Terriza2007} and electrons \cite{McMorran2011,Guzzinati2014,Grillo2014}. Furthermore extrinsic OAM states have also been observed in neutrons, using spiral phase plates \cite{Clark2015} and magnetic gradients \cite{Sarenac2019}. In the latter case spin-orbit coupled states are generated \cite{Sarenac2018}. It has also been demonstrated that magnetic quadrupoles can generate spin-orbit states in neutral spin 1/2 particles \cite{Hinds2000,Nsofini2016}. The aforementioned methods require a beam with exceptional collimation (0.01$^{\degree}$-0.1$^{\degree}$ divergence) if intrinsic OAM is the goal. Furthermore the incident particles must be on the optical axis. These two requirements limit the available flux to an impractical level. For this reason intrinsic OAM has not been observed in neutrons to date \cite{Cappelletti2018}. The additional quantum degree of freedom offered by OAM provides utility in the realm of quantum information \cite{Vallone2014,Fickler2014,Ding2019}. Additionally in neutrons the additional degree of freedom may help improve existing tests of quantum contextuality \cite{Hasegawa2010,Shen2020}. Furthermore neutrons carrying net OAM may reveal additional information on atomic nuclei in scattering experiments \cite{Afanasev2019}. \par
In this paper we propose a method by which intrinsic spin-orbit states can be generated in an arbitrarily collimated beam of neutral spin 1/2 particles. This removes flux limitations and allows for the construction of spin-orbit optical equipment for neutrons. We show that a static homogeneous electric field polarized along the direction of particle propagation induces longitudinal spin-orbit states, while a transversely polarized electric field generates transverse spin-orbit states. The latter type of OAM has not yet been observed in massive free particles. Furthermore we confirm previous results that the total angular momentum of a particle is conserved in static electric fields \cite{Bruce2020}. As shown by Schwinger \cite{Schwinger1948} in an electric field the particle spin couples to the cross product between the electric field strength and the particle momentum. Phase shifts due to this coupling have been observed in Schwinger scattering \cite{Shull63a,Shull63b,Voronin2000,Gentile2019}, the Aharonov Casher effect \cite{Aharonov1984,Cimmino1989,Cimmino2000} and in measurements of the neutron electric dipole moment \cite{Dress1977,Harris1999,Fedorov2010,Piegsa2013} where it can be a major systematic effect. In dynamical diffraction from non-centroysmmetric crystals spin rotations of up to 90$^{\degree}$ have been observed \cite{Voronin2000,Fedorov2010}, due to large interplanar fields. Recently the Schwinger coupling has been used to image electric fields with polarized neutrons \cite{Jau2020}. However to date no tests for OAM have been conducted.
 \par
\section{Theoretical Framework.} An observer moving through an electric field, $E$, will experience a magnetic field $B'$. In the low velocity limit when $v<<c$ the magnetic field can be written as \cite{Zangwill}
\begin{equation}\label{Lorentz}
	\vec{B}'=\vec{v}\times \frac{\vec{E}}{c^2}
\end{equation}
Inversely in the lab frame a moving magnetic moment will appear to have a small electric dipole moment $\vec{d}'=\frac{\vec{v}\times\vec{\mu}}{c^2}$. Hence a spin 1/2 particle with magnetic moment $\vec{\mu}$ experiences a Zeeman shift $\vec{d}'\cdot\vec{E}=\vec{\mu}\cdot\vec{B}'$ when moving through an electric field. Therefore the Schroedinger equation is 
\begin{equation}\label{Schoed}
	[-\nabla^2-\frac{\gamma}{c^2}\vec{\sigma}\cdot(\vec{p}\times\vec{E})]\psi=\epsilon\psi
\end{equation}
with $\gamma$ the gyromagnetic ratio and $\vec{\sigma}$ the Pauli matrices. The wavefunction is described by a spinor $\psi=\begin{pmatrix}
\psi_+(x,y,z) \\
\psi_-(x,y,z)
\end{pmatrix}$, where the index $\pm$ refers to the spin state parallel or anti-parallel to the z-axis respectively.
\par
\subsection{Transmission Geometry - Longitudinal OAM.} 

First we will consider the longitudinal spin-orbit effect. We will assume that the extent of the electric field is semi-infinite and that it is parallel to the z-axis. Hence the Schroedinger equation can be written as
\begin{equation}\label{Eq2}
	-\nabla^2\psi_\pm+iC(\frac{\partial}{\partial y}\pm i\frac{\partial}{\partial x})\psi_\mp =\epsilon\psi_\pm
\end{equation} \\
with $C=\frac{\gamma E_z}{c^2}$. The incident wave will be described by $\psi^I_\pm=f(r,\phi)e^{-ikz}$. Note that for a non-zero coupling this effect requires the incident wavefunction to have a transverse momentum component. By applying a Fourier transform over the x and y coordinates the PDE (Eq. \ref{Eq2}) is simplified to a coupled second order ODE.
\begin{equation}\label{FourierCoupledODE}
-(\frac{\partial^2}{\partial z^2}-k_r^2+\epsilon)\hat{\psi}_\pm \mp iCk_r e^{\mp i\phi } \hat{\psi}_\mp=0
\end{equation}
Here we have also transformed the equation to cylindrical coordinates with $k_r^2=k_x^2+k_y^2$ and $k_x\pm ik_y=k_re^{\pm i\phi}$. It is noteworthy that in the spectral domain the potential, $C(k_x\sigma_y+k_y\sigma_x)$, closely resembles that of the magnetic quadrupole in real space. This gives an intuitive reason as to why a static electric field mimics the action of a quadrupole in reciprocal space. Hence an electric field is more effective for large divergences (i.e. large $k_r$). We diagonalize equation \ref{FourierCoupledODE}, by applying a transformation of the form $\hat{\psi}=T\hat{\psi}'$ and multiplying the Hamiltonian by $T^{-1}$ from the left.
\begin{equation}\label{FourierODE}
[-(\frac{\partial^2}{\partial z^2}-k_r^2+\epsilon)\mp C k_r]\hat{\psi}'_\pm=0
\end{equation}
For this particular diagonalization $T$ is given by $\begin{pmatrix}
&ie^{-i\phi} \ \ &-ie^{-i\phi} \\
&1 \ \ &1
\end{pmatrix}$. The general solution to Eq. \ref{FourierODE} is simply a superposition of a forward and backward propagating plane wave for each spin state 
\begin{equation}\label{TransformedSolution}
	\hat{\psi}'=\begin{pmatrix}
	\hat{t}_1e^{ik_{+}z}+\hat{t}_2e^{-ik_{+}z} \\
	\hat{t}_3e^{ik_{-}z}+\hat{t}_4e^{-ik_{-}z}
	\end{pmatrix}
\end{equation}
with $k_{\pm}=\sqrt{\epsilon-k_r^2\pm Ck_r}$. Amplitudes of the backward propagating solutions, $\hat{t}_1$ and $\hat{t}_3$, are zero. The general solution for $\hat{\psi}$ is simply found by applying the transformation $T\hat{\psi}'$.
\begin{equation} \label{GeneralFourierSolution}
	\hat{\psi}=\begin{pmatrix}
	ie^{-i\phi}[\hat{t}_2e^{-ik_{+}z}-\hat{t}_4e^{-ik_{-}z}]\\
	\hat{t}_2e^{-ik_{+}z}+\hat{t}_4e^{-ik_{-}z}
	\end{pmatrix}
\end{equation}
To determine the values of $\hat{t}_2$, $\hat{t}_4$ and the reflection coefficients $\hat{r}_\pm$ we apply the boundary conditions 
\begin{equation}\label{Boundary}
\begin{aligned}
	&\hat{\psi}(k_r,\phi,z=0)=\hat{f}_\pm+\hat{r}_\pm \\
	&\hat{\psi}_z(k_r,\phi,z=0)=ik_z(\hat{r}_\pm-\hat{f}_\pm)
\end{aligned}
\end{equation}
Here the subscript $z$ under $\psi$ denotes the partial derivative to the $z$ coordinate. $\hat{f}_\pm(k_r,\phi)$ denotes the 2D Fourier transform of the incident wavefunction. This boundary value problem can be formulated as the following matrix vector problem
\begin{equation}\label{BVPMatrix}
	\begin{pmatrix}
	&1 \ \ &-1 \ \ &1 \ \ &0 \\
	&1 \ \ &1 \ \ &0 \ \ &-1 \\
	&k_{+} \ \ &-k_{-} \ \ &-k_z \ \ &0 \\
	&k_{+} \ \ &k_{-} \ \ &0 \ \ &k_z
	\end{pmatrix}
	\begin{pmatrix}
	\hat{t}_2 \\
	\hat{t}_4 \\
	i\hat{r}_+e^{i\phi} \\
	\hat{r}_-\\
	\end{pmatrix}= \begin{pmatrix}
	-i\hat{f}_+e^{i\phi} \\
	\hat{f}_- \\
	-ik_z\hat{f}_+e^{i\phi} \\
	k_z\hat{f}_- \\
	\end{pmatrix}
\end{equation}
By inverting the above 4x4 matrix we find the transmission and reflection coefficients
\begin{equation}\label{T}
	\begin{aligned}
	&\hat{t}_{\binom{2}{4}}=\frac{\mp ik_z\hat{f}_+e^{i\phi}+k_z\hat{f}_-}{(k_z+k_\pm)}\\
	&\hat{r}_\pm=\pm \frac{(k_z^2-k_+k_-)\hat{f}_\pm \mp ik_z(k_+-k_-)e^{\mp i\phi}\hat{f}_\mp }{(k_++k_z)(k_-+k_z)}\\
	\end{aligned}
\end{equation}
which leads us to the solution for the transmitted waves
\begin{equation}\label{transmitted waves}
	\hat{\psi}_\pm=
	\frac{k_z\hat{f}_\pm\pm ik_z\hat{f}_\mp e^{\mp i\phi}}{(k_z+k_+)}e^{-ik_+z}+\frac{k_z\hat{f}_\pm\mp ik_z\hat{f}_\mp e^{\mp i\phi}}{(k_z+k_-)}e^{-ik_-z}
\end{equation}
Looking at this expression we can see that the total angular momentum $J=S+L$ of the wave is conserved in a static electric field, since a spin flip is compensated by a change in OAM. \par $\hat{f}_\pm(k_r,\phi)$ can be expanded such that $\hat{f}_\pm(k_r,\phi)=\sum_\ell \hat{f}^\ell_\pm(k_r)e^{i\ell\phi}$, where $\hat{f}_\pm^l(k_r)$ is given by the azimuthal Fourier Transform
\begin{equation}\label{FourierAzimuth}
	\hat{f}_\pm^\ell=\int_{0}^{2\pi} \hat{f}_\pm(k_r,\phi)e^{-i\ell\phi}d\phi
\end{equation}
The solution in real space can be obtained by applying the Bessel/Hankel transform to Eq. \ref{transmitted waves}.
\begin{equation}\label{InverseTransformation}
\begin{aligned}
	\psi_\pm=&\sum_\ell i^{-\ell}k_z e^{i\ell\theta}\\\int_0^{\infty}&
	\frac{\hat{f}^\ell_\pm\pm i\hat{f}^{\ell\pm1}_\mp}{(k_z+k_+)}e^{-ik_+z}+\frac{\hat{f}^\ell_\pm\mp i\hat{f}^{\ell\pm1}_\mp}{(k_z+k_-)}e^{-ik_-z}J_\ell(k_r r)k_rdk_r
\end{aligned}
\end{equation}
It is instructive to look at the solution of Eq. \ref{InverseTransformation} for an incident wavefield, $\psi^I_{\pm}$, described by a Bessel beam carrying no OAM, $\psi^I_{\pm}=b_\pm J(k_\rho r)e^{-ik_z z}$, with $b_\pm$ the amplitude of the up and down spin state respectively and $k_\rho$ the transverse momentum component of the incident wave. Hence $\hat{f}_\pm^{\ell\neq 0}=0$ and $\hat{f}_\pm^0(k_r)=b_\pm\frac{\delta(k_r-k_\rho)}{k_r}$ with $\epsilon=k_z^2+k_\rho^2$. In this case the solution is trivial
\small
\begin{equation}\label{solved}
\begin{aligned}
	&\psi_\pm^0=k_zb_\pm J_0(k_\rho r)
	(\frac{e^{-i\sqrt{k_z^2+Ck_\rho}z}}{(k_z+\sqrt{k_z^2+Ck_\rho})}+\frac{e^{-i\sqrt{k_z^2-Ck_\rho}z}}{(k_z+\sqrt{k_z^2-Ck_\rho})})\\&\psi_\pm^{1}=\pm k_zb_\mp J_1(k_\rho r)
	(\frac{e^{-i\sqrt{k_z^2+Ck_\rho}z}}{(k_z+\sqrt{k_z^2+Ck_\rho})}-\frac{e^{-i\sqrt{k_z^2-Ck_\rho}z}}{(k_z+\sqrt{k_z^2-Ck_\rho})})
\end{aligned}
\end{equation}
\normalsize
where $\psi_\pm^0$ and $\psi_\pm^1$ are the components with and without OAM respectively, such that $\psi_\pm=\psi_\pm^0+e^{\mp i\theta}\psi_\pm^1$. For a collimated beam geometry we may use $k_\rho=k_z\tan(\alpha)\approx k_z\alpha$, where $\alpha$ is the beam divergence. Furthermore if $Ck_\rho$ is sufficiently small we may linearize the square root terms in equation \ref{solved} and obtain a much simpler expression for the wavefunction.
\begin{equation}\label{linear}
\begin{aligned}
	&\psi_\pm= [b_\pm\cos(\frac{\gamma E_z\alpha}{2c^2}z)J_0(k_\rho r)\\&\pm b_\mp \sin(\frac{\gamma E_z\alpha}{2c^2}z)e^{\mp i\theta}J_1(k_\rho r)]e^{-ik_zz}
	\end{aligned}
\end{equation}
\begin{figure*}
	\includegraphics[width=18cm]{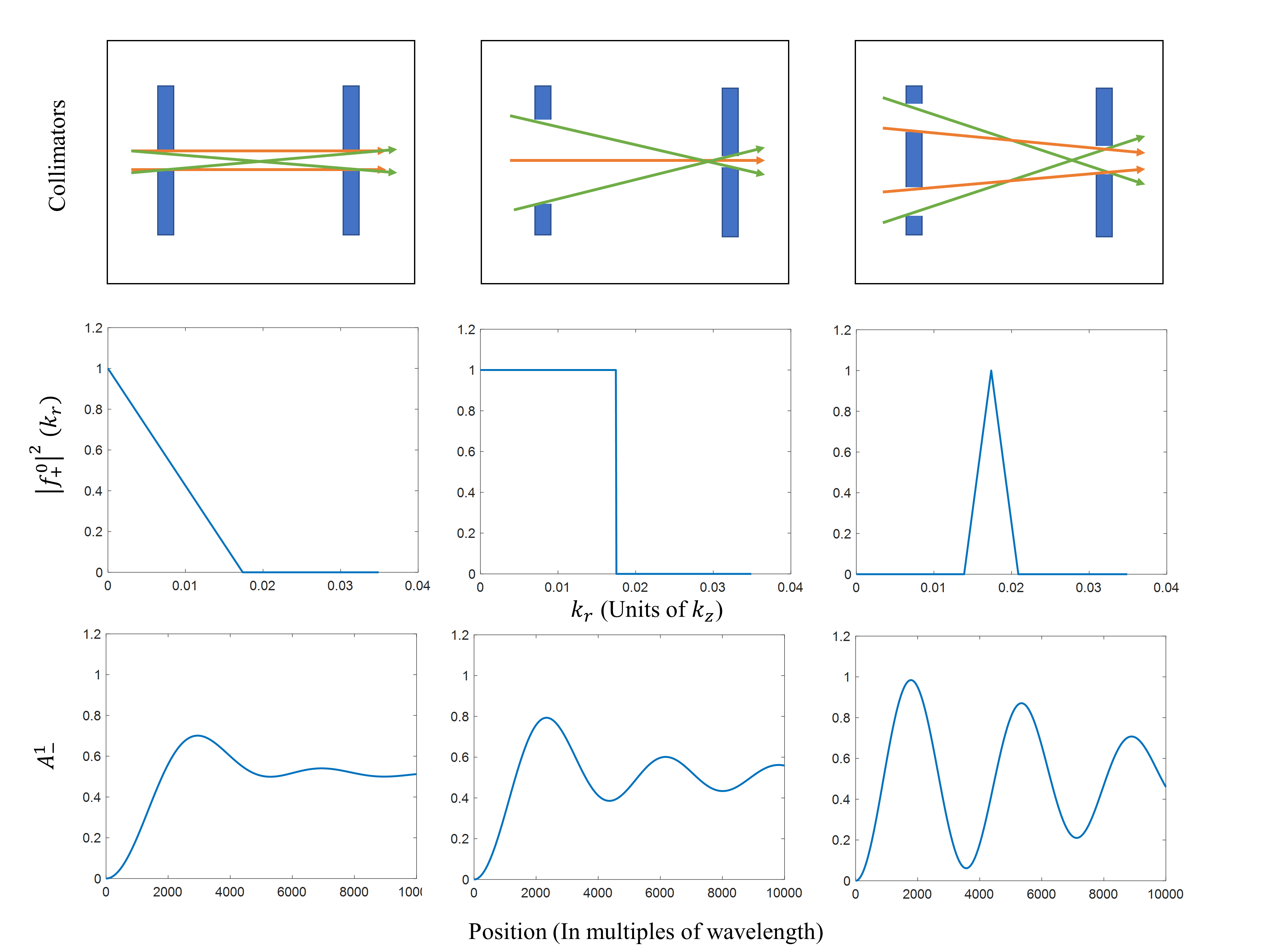}
	\caption{For various common collimator types such as two identical pinholes (left), a large exit and a comparatively small pinhole (middle) and an annulus with pinhole (right), we show the possible beam paths through a hypothetical instrument (top) and the respective divergence profiles (middle). The paths with the lowest divergence are drawn in orange, while the maximum divergence paths are shown in green. The divergence profiles are used as $|f_+|^2(k_r)$ in equation \ref{Superpostion} to determine the probability of finding the particle in the $l=1$ OAM state as a function of the $z$ position in an electric field (bottom). The parameters are chosen such that $k_z=1$, $\epsilon\approx k_z^2$ and $C=0.1$. We note that that in a real instrument these divergence profiles might represent the incoherent average of all possible incident wavefields and not the actual transverse incident wavefield of a single neutron.}
	\label{Distributions}
\end{figure*}
A longitudinal beam twister device may be constructed using a parallel plate capacitor, with the surfaces of the plates normal to the beam. The voltage required to fully twist the beam from the $\ell=0$ state into the $\ell=\pm 1$ state is given by
\begin{equation}\label{Voltage}
	V=\frac{\pi c^2}{\gamma \alpha}
\end{equation}

These equations are valid for single Bessel beams. However Bessel functions are not normalizable \cite{Bliokh2017} and therefore have infinite coherence, making them unphysical. In a realistic setup we always have a normalizable superposition of Bessel beams, which have finite coherence. This superposition interferes and results in damping of spin orbit production, due to dephasing. This interference can be described by solving Eq. \ref{InverseTransformation} for an arbitrary divergence profile. 
Though we can also determine the probability of the particle being in the mth OAM state as a function of $z$ without the inverse transform, Eq. \ref{InverseTransformation}, by simply calculating the projection of Eq. \ref{transmitted waves} on $e^{im\phi}$ and integrating the absolute value squared of this expression over $k_r$:
\begin{equation}\label{Superpostion}
\begin{aligned}
	A^m_\pm=\int |\hat{\psi}_\pm^m|^2 k_r dk_r=\int|\psi_\pm^m|^2rdr
\end{aligned}
\end{equation}
 with $\hat{\psi}^m=<e^{im\phi}|\hat{\psi}>$, the azimuthal Fourier transform (eq. \ref{FourierAzimuth}) of $\hat{\psi}$. Here we have also used Parsevals theorem to demonstrate that the value of $A^m$ is the same in real and reciprocal space. Solutions of equation \ref{Superpostion} for the most common divergence profiles, $|f|^2(k_r,\phi)$ are shown in figure \ref{Distributions}. Here we see dephasing effects which causes the contrast of $A^1_-$ to wash out as the wave penetrates deeper into the electric field. As the transverse wavelength spread is decreased the dephasing effects are also reduced. This is analogous to dephasing seen in magnetic spin echo instruments, due to the longitudinal wavelength spread \cite{Mezei1980}. \par
 Equation \ref{Voltage} demonstrates that for particles with a divergence of $1^{\degree}$ propagating through a capacitor we require a voltage drop of $88.4GV$ to put a neutron into an OAM state with $\ell=\pm 1$. Obviously this is not feasible. For colder particles it is possible to use zone plates which consist of concentric rings of periodically spaced absorber material to increase the transverse momentum, $k_r$, thereby decreasing the required voltage drop. Such Fresnel lenses have been produced for the purpose of imaging with very cold neutrons \cite{Kearney1980}.
 \par 
\subsection{Reflection Geometry - Quasi Transverse OAM.} 
\begin{figure}
	\includegraphics[width=9cm]{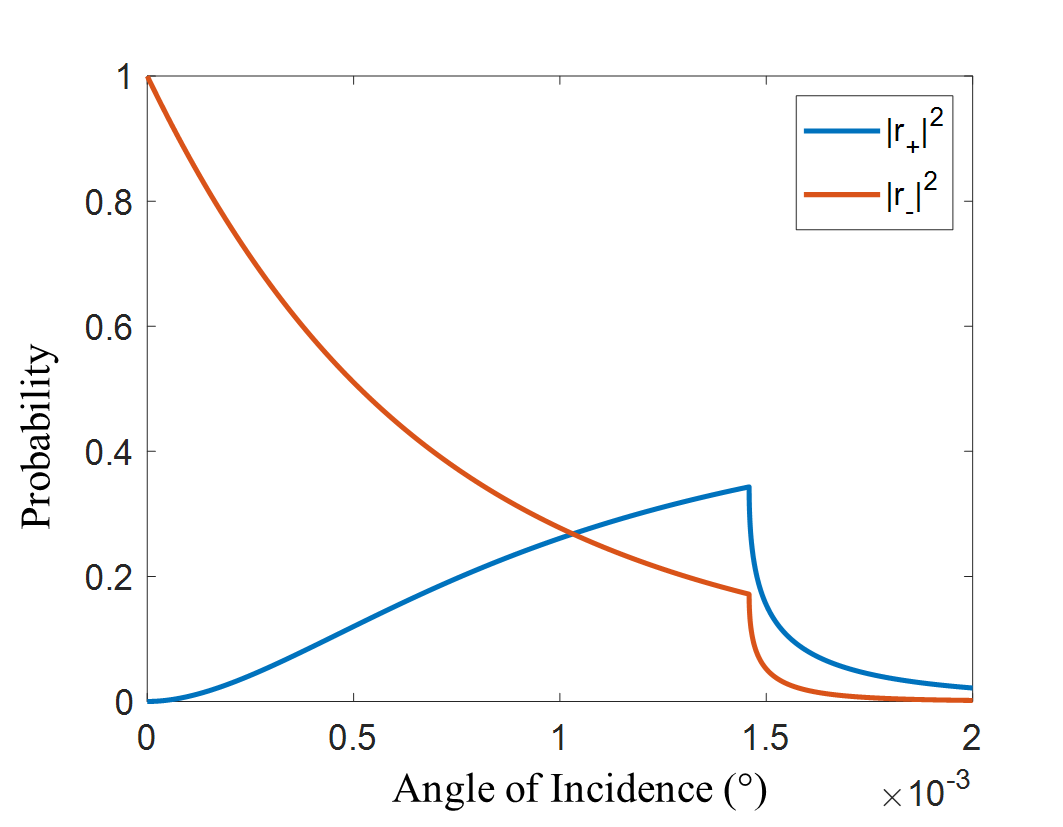}
	\caption{Reflection probability according to equation \ref{T}, $\hat{f}_+=0$ and $\hat{f}_-=1$. A wavelength of $2$ \AA \ and an electric field of $10^{10} V/m$ are assumed. The blue curve corresponds to a spin flip reflection which generates OAM, while the red curve shows the non spin flip reflection probability.}
	\label{Reflection2}
\end{figure}

 Next we consider waves interacting with an electric field interface at grazing incidence angles. This results in a more pronounced coupling, due to a larger $k_r$ and a smaller value for $k_z$. The OAM carried by the transmitted and reflected waves in this case is quasi-transverse to the wavevector $\vec{k}$. Since the quantization axis of the OAM is normal to interface, the incident wave must be described by an infinite superposition of OAM modes. Nonetheless the mean OAM of the transmitted and reflected waves can be raised or lowered by one unit of $\hbar$ with respect to the incident OAM. The reflection probability $|r_\pm|^2$ as a function of incident angle is shown in Fig. \ref{Reflection2}, for an electric field of $10^{10} V/m$ (found in electric double layers \cite{Toney1995,Ferechmin2002}), a neutron wavelength of $2$ \AA \ and an initial spin aligned along the $-z$ direction. We can deduce that the optimal angle of reflection is around $0.001^{\degree}$. Hence this method of OAM generation is likely not feasible due to flux limitations.
\par 
\subsection{Transmission Geometry - Transverse OAM}

The flux limitations can be overcome by considering transmission through a transversely polarized electric field which leads to the generation of transverse spin-orbit states. To demonstrate this we consider the time dependent Schroedinger equation for a neutral spin 1/2 particle in an electric field
\begin{equation}\label{SchoedTime}
[-\nabla^2-\frac{\gamma}{c^2}\vec{\sigma}\cdot(\vec{p}\times\vec{E})]\psi=-i\frac{\partial}{\partial t}\psi
\end{equation}
Again we will assume that the electric field is polarized along the z-direction. However this time we will consider a field which extends infinitely in space. 
To reduce the problem to an ordinary differential equation we apply an unbounded Fourier transform to the spatial coordinates. In cylindrical coordinates this leads to 
\begin{equation}\label{SchoedTimeFourier}
\epsilon\hat{\psi}_\pm\mp iCk_r e^{\mp i\phi } \hat{\psi}_\mp=-i\frac{\partial}{\partial t}\hat{\psi}_\pm
\end{equation}
$\epsilon$ now denotes the kinetic energy parameter $k_r^2+k_z^2$. Once again we diagonalize this set of equations using the transform $\hat{\psi}=T\hat{\psi}'$
\begin{equation}\label{SchoedTimeFourDiag}
[\epsilon\mp Ck_r]\hat{\psi}_\pm'=-i\frac{\partial}{\partial t}\hat{\psi}_\pm'
\end{equation}
Applying the initial conditions $\hat{\psi}_\pm(t=0)=\hat{a}_\pm(k_r,\phi,k_z)$ we can determine the homogeneous solution of equation \ref{SchoedTimeFourier}. 
\begin{equation}\label{TimeSolution}
	\hat{\psi}_\pm=e^{i\epsilon t}[a_\pm \cos(Ck_r t)\pm a_\mp \sin(Ck_rt)e^{\mp i\phi}]
\end{equation}
which appears almost equivalent to equation \ref{linear}. If the wave propagates along the y-direction the value of $k_r$, which may be approximated by $k_y$ is a factor $10^2-10^3$ larger than in the longitudinal case (equation \ref{linear}). Hence the required electric field integral to raise or lower the mean OAM is reduced to a more practical level. The incident wave in this case must be described by an infinite superposition of transverse OAM modes. Upon being transmitted through an ideal beam twister device the mean $\ell$ value of this superposition will be raised or lowered by one. In this paper we assume that $\hat{a}_\pm$ can be approximated by a Gaussian model. The standard deviation in $k_x$ direction can be expressed in terms of a symmetry factor $R$ and the standard deviation in $k_y$ direction $\sigma_y$: $\sigma_x=R\sigma_y$. Such that $\hat{a}_\pm=e^{-\frac{(k_y-k_y').^2}{\sigma_y^2}}e^{-\frac{k_x^2}{R^2\sigma_y^2}}$, with $k_y'$, the mean momentum in the y-direction. This Gaussian can be expanded in its various OAM components by means of the azimuthal Fourier transform.
\begin{figure}
	\includegraphics[width=9cm]{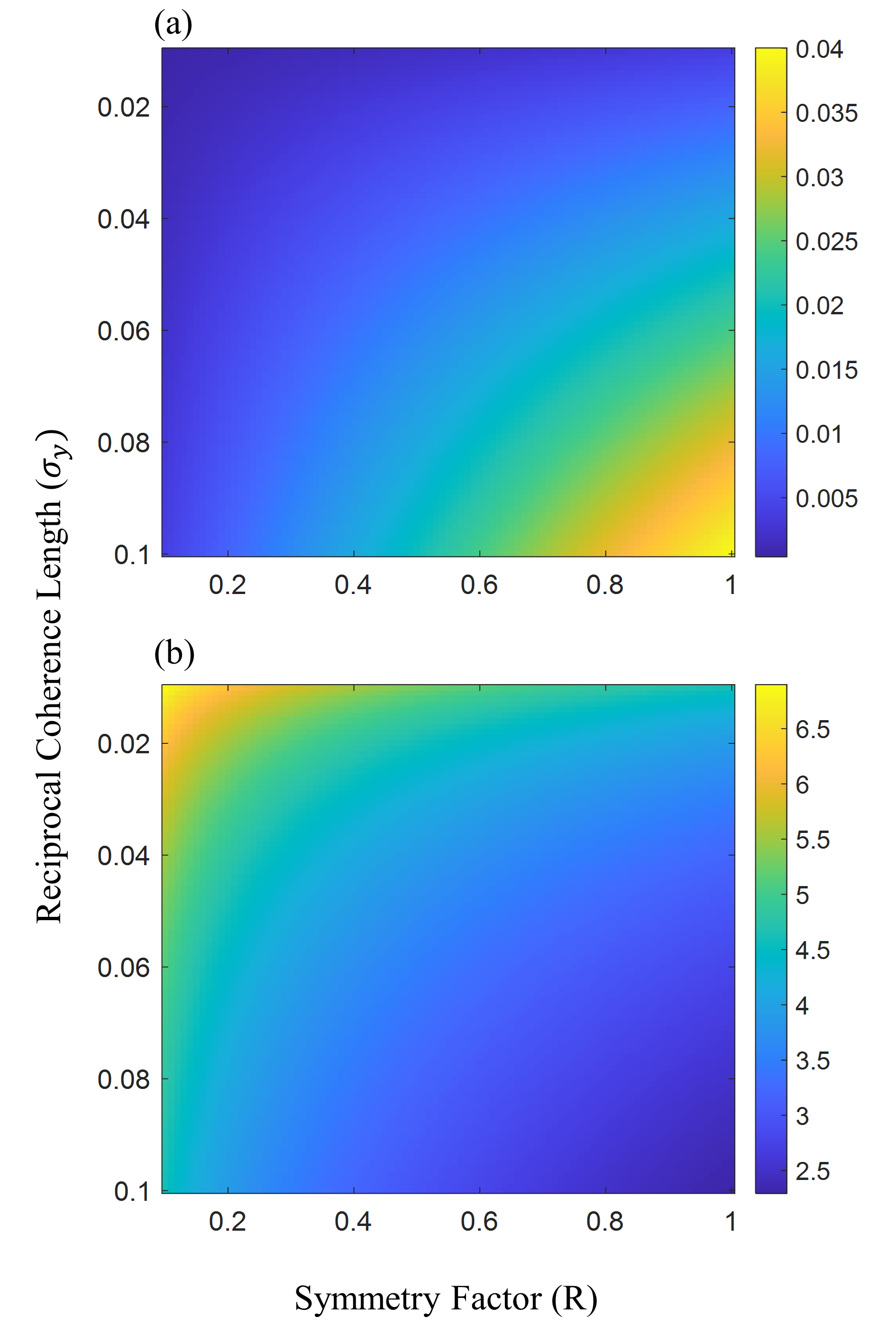}
	\caption{(a) The amplitude of the first OAM mode $A^1$ and (b) the logarithm of the OAM bandwidth $\sigma_\ell$ of a twisted Gaussian wavepacket plotted as a function of the reciprocal coherence length, $\sigma_y$ and the symmetry factor $R$, assuming $k_y'=1$.}
	\label{Amplitudes}
\end{figure}
Upon passing through an appropriate electric field the index $\ell$ is raised or lowered by 1. Using this and equation \ref{Superpostion} the amplitude of the $\ell=1$ OAM mode, $A^1$, can be calculated. We may also define an OAM bandwidth in terms of the standard deviation
\begin{equation}\label{Bandwidth}
	\sigma_\ell=\sqrt{<L_z^2>-<L_z>^2}
\end{equation}
with $<L_z>=\sum_\ell \ell A^\ell$ and $<L_z^2>=\sum_\ell \ell^2 A^\ell$. Both the OAM amplitude $A^1$ and the OAM bandwidth, $\sigma_\ell$, are shown as a function of the reciprocal longitudinal coherence length $\sigma_y$ and the symmetry factor $R$ in Fig. \ref{Amplitudes}. One can see that a small coherence length (large $\sigma_y$) leads to a larger amplitude, $A^1$ and a tighter bandwidth, $\sigma_\ell$. Analogously a large symmetry factor $R$ corresponds (i.e. a large beam divergence) to a larger amplitude, $A_1$ and a small bandwidth, $\sigma_\ell$. \par
\begin{figure}
	\includegraphics[width=9cm]{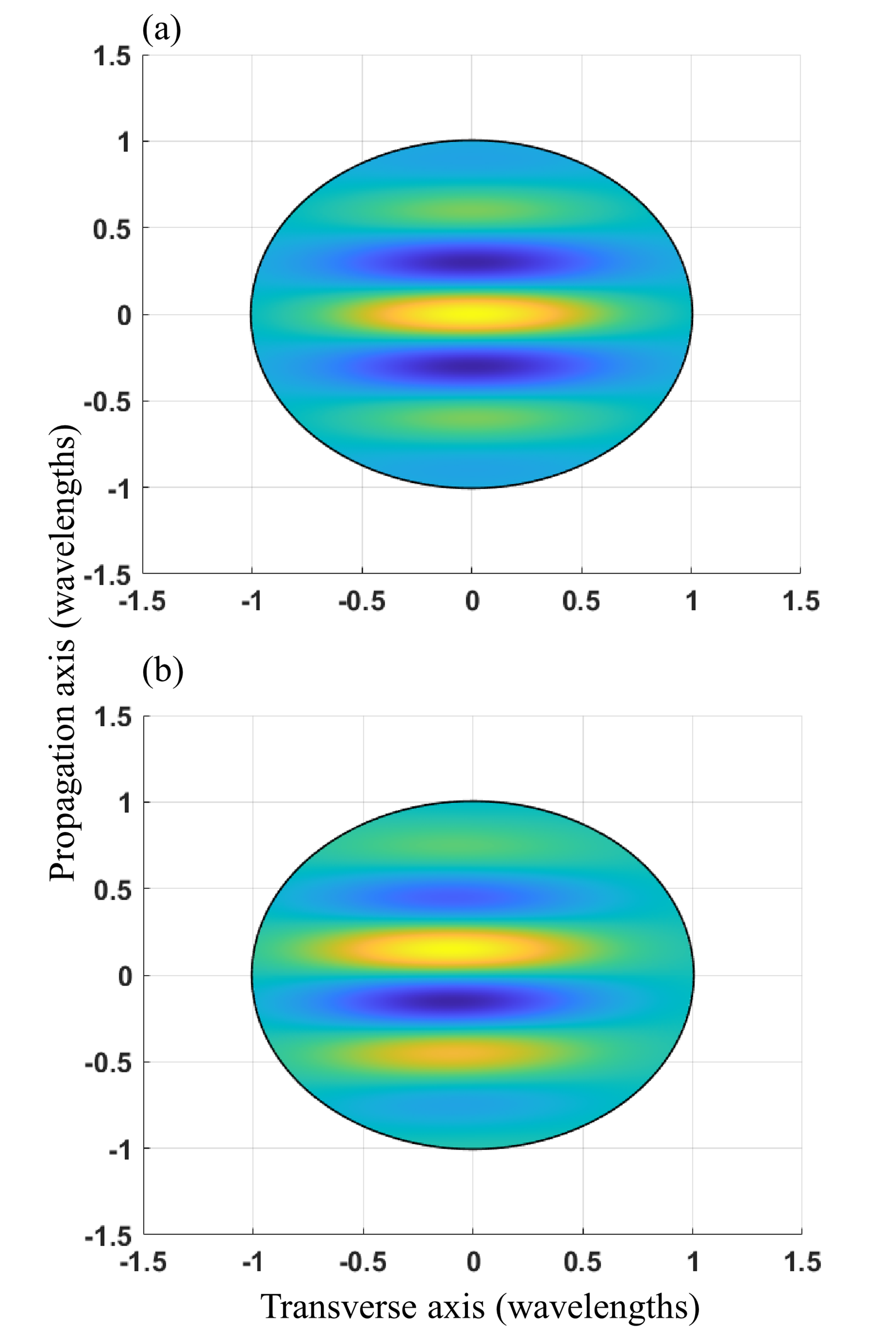}
	\caption{Surface plots of the real parts of Gaussian wavepackets in real space, with $k_y'=1$, $\sigma_y^2=0.1$ and $R=1$ carrying (a) no orbital angular momentum and (b) one unit of transverse orbital angular momentum.}
	\label{TOAM}
\end{figure}

In Fig. \ref{TOAM} we show one such Gaussian wavepacket carrying transverse OAM in real space. The wavepacket with OAM appears to be displaced along the transverse axis, while along the longitudinal axis the wavepacket is shifted by $\pi/2$.
\begin{figure*}
	\includegraphics[width=18cm]{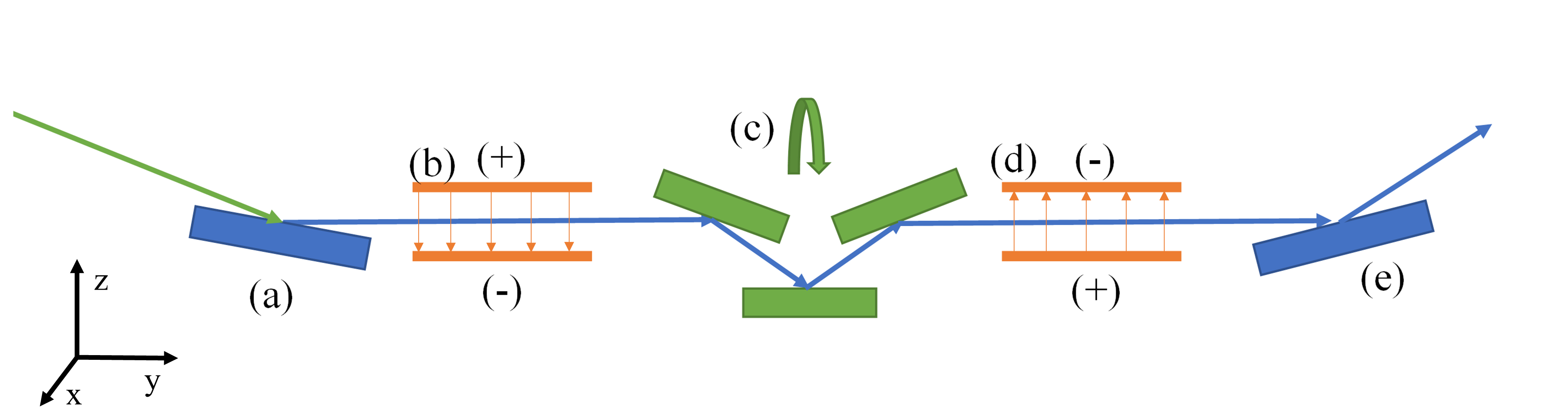}
	\caption{Simplified schematic of the proposed proof of concept experiment to demonstrate the generation of transverse OAM by electric fields. The setup consists of a polarizing supermirror (a), an electric beam twister (b), a set of non polarizing mirrors which can be rotated around a common beam axis (c),  another beam twister with opposite polarity (d) and a supermirror for spin polarization analysis (e).}
	\label{Setup}
\end{figure*} 

\section{Proposed Methodology.} Based on the previous theoretical
 analysis we propose a proof of concept experiment with neutrons to demonstrate that magnetic neutral spin 1/2 particles can obtain quanta of transverse OAM when traversing an electric field polarized perpendicular to the flight direction.
 The beam twister device will consist of a one meter long evacuated flight tube loaded with two electrodes 1 mm apart. A voltage is applied across the electrodes to generate the experimentally highest possible field in a high vacuum environment ($10^7-10^8 \ V/m$). Such a beam twister can generate an OAM carrying wave with an amplitude between 2\% and 20\%. To measure the OAM we propose an experiment similar to \cite{Leach2002}, which was designed for photons. The experimental setup would employ two supermirrors to spin polarize and analyze the beam, two beam twisters to generate and analyze spin-orbit coupling and a set of three mirrors in between the two beam twisters as a means of rotating the image and inverting the OAM quantum number. This image rotation implies that the quantization axis of the transverse OAM is rotated around the propagation axis. If the dove prism is positioned such that the OAM is flipped the second beam twister will fail to properly decouple the spin-orbit states, thereby leading to destructive interference at the detector. On the other hand the prism may also be rotated into a position which does not alter the OAM. In this case the second beam twister successfully decouples the neutron spin-orbit states and constructive interference is seen at the detector. Hence by rotating the dove prism at a constant frequency a time dependent modulation will be seen in the neutron intensity. Since the effects of all components described in this setup are wavelength independent, the experiment can exploit the high thermal flux of a white neutron beam. The proposed setup is shown in Fig. \ref{Setup}.

\par
\section{Conclusion.} We have provided a theoretical framework which predicts that magnetic neutral spin 1/2 particles propagating through a static electric field acquire OAM parallel to the electric field axis. Furthermore we have illustrated a proof of concept experiment which could verify the generation of transverse OAM in neutrons transmitted through an electric fields.
\par
\begin{acknowledgments}
The authors thank Victor de Haan for fruitful discussion. Furthermore we would like to extend our gratitude to Andrei Afanasev for checking the mathematical derivations in this paper. This work was financed by the Austrian Science Fund (FWF), Project No. P30677 and P34239.
\end{acknowledgments}

\end{document}